\documentclass[3p]{elsarticle}

\usepackage{lineno,hyperref,amsmath,amssymb}
\modulolinenumbers[5]

\journal{Phys. Lett. B}

\newcommand{\amc}{{\sc MadGraph5\_aMC@NLO}}
\newcommand{\ml}{{\sc MadLoop}}

\newcommand{\nlo}{{\sc NLOCT}}
\newcommand{\gev}{\,\textrm{GeV}}

\newcommand{\alphas}{\alpha_s}
\newcommand{\hwbb}{pp\to H^{\pm} W^{\mp} b \bar b}

\begin{document}

\begin{frontmatter}

\title{Accurate predictions for charged Higgs production: \\closing the $m_{H^{\pm}}\sim m_t$ window}

\author{ C\'{e}line Degrande}
\address{ Institute for Particle Physics Phenomenology, Department of Physics,\\ 
Durham University, Durham DH1 3LE, United Kingdom} 
\ead{celine.degrande@durham.ac.uk}
\author{Rikkert Frederix}%
\address{ Technische Universit\"at M\"unchen,\\
James-Franck-Str.~1, D-85748 Garching, Germany}
\ead{rikkert.frederix@tum.de}
\author{Valentin Hirschi}%
\address{SLAC, National Accelerator Laboratory,\\
    2575 Sand Hill Road, Menlo Park, CA 94025-7090, USA}
\ead{vahirsch@slac.stanford.edu}
\author{Maria Ubiali}%
\address{ Cavendish Laboratory, HEP group,\\
University of Cambridge, J.J.~Thomson Avenue, Cambridge CB3 0HE, United Kingdom}
\ead{ubiali@hep.phy.cam.ac.uk}
\author{Marius Wiesemann}%
\address{Physik-Institut, Universit\"at Z\"urich,\\ 
Winterthurerstr.~190, 8057 Zurich, Switzerland }
\ead{mariusw@physik.uzh.ch}
\author{Marco Zaro}%
\address{Sorbonne Universit\'es, UPMC Univ.~Paris 06,\\ 
UMR 7589, LPTHE, F-75005, Paris, France
\\and \\ CNRS, UMR 7589, LPTHE, F-75005, Paris, France}
\ead{zaro@lpthe.jussieu.fr}

\begin{abstract}
We present predictions for the total cross section for the production of a charged
Higgs boson in the
intermediate-mass range ($m_{H^{\pm}}\sim m_t$) at the LHC, focusing on a type-II two-Higgs-doublet model. Results are obtained
at next-to-leading
order (NLO) accuracy in QCD perturbation theory, by studying the full process $pp\to H^\pm
W^\mp b \bar b$ in the complex-(top)-mass scheme with massive bottom quarks. Compared to lowest-order predictions, 
NLO corrections have a sizeable impact: they increase the cross section by roughly 50\% and 
 reduce uncertainties
due to scale variations by more than a factor of two. Our computation
reliably interpolates between the low- and high-mass regime.
Our results provide the first NLO prediction for charged Higgs production in the intermediate-mass range and therefore 
 allow to have NLO accurate predictions in the full $m_{H^{\pm}}$ range. 
The extension of our results to different realisations of the two-Higgs-doublet model or to the supersymmetric case is also discussed. \end{abstract}

\begin{keyword}
Charged Higgs \sep Resonant diagrams \sep Complex Mass Scheme \sep  Top Quark
\sep Two-Higgs-Doublet Models
\end{keyword}

\end{frontmatter}

%\linenumbers

Charged Higgs bosons appear in the scalar sector of several Standard Model (SM) extensions, and
are the object of various beyond the Standard Model (BSM) searches at the LHC. 
As the SM does not include any elementary charged scalar particle, the observation of a 
charged Higgs boson would necessarily point to a non-trivially extended scalar sector.

In this paper we focus on a generic two-Higgs-doublet model (2HDM), which is one of the simplest SM extensions 
featuring a charged scalar. Within this class of models, two isospin doublets are introduced
to break the $SU(2)\times U(1)$ symmetry, leading to the existence of five physical Higgs 
bosons, two of which are charged particles ($H^{\pm}$). 
Imposing flavour conservation, there are four possible ways 
to couple the SM fermions to the two Higgs doublets~\cite{Branco:2011iw}. Each of the four ways
gives rise to rather different phenomenologies. In this work, we consider the so-called type-II 2HDM 
(although we will discuss how our results can be generalised to other types), 
in which one doublet couples to up-type quarks 
and the other to down-type quarks and charged leptons.
%Among others, the Minimal Supersymmetric Standard Model (MSSM), up to SUSY corrections, belongs to the
%type-II 2HDM category, where the first Higgs doublet gives mass to the up-type fermion and the second to the down-type fermions.

\begin{figure}[h]
    \centering
    \begin{tabular}{ccc}
    \includegraphics[clip=true, height=1.6cm]{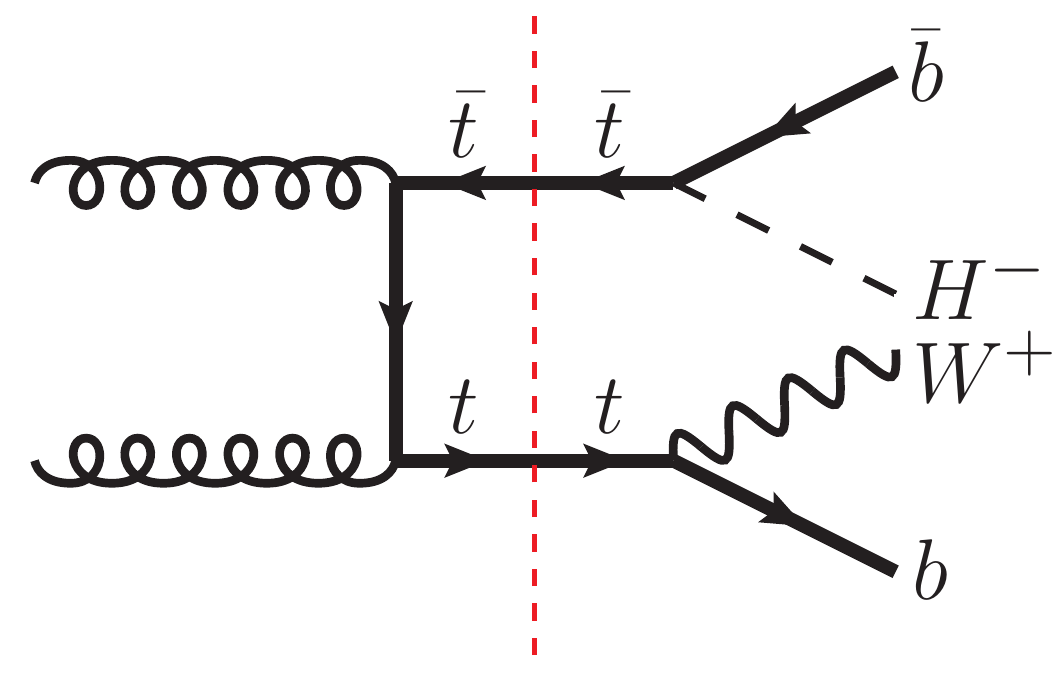} &&
    \includegraphics[clip=true, height=1.6cm]{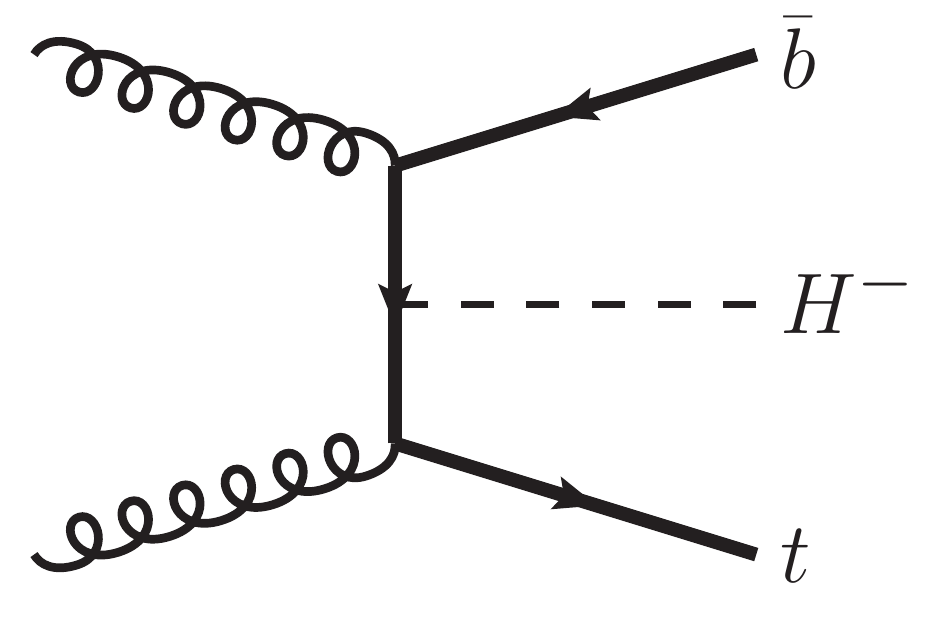}\\
     (a) && (b)
     \end{tabular}
    \caption{\label{fig:diaglowhighmass} Sample LO diagrams for (a) light and (b) heavy charged Higgs production.}
\end{figure}

\begin{figure*}[h]
    \centering
    \begin{tabular}{ccccccc}
    \includegraphics[clip=true, height=1.6cm]{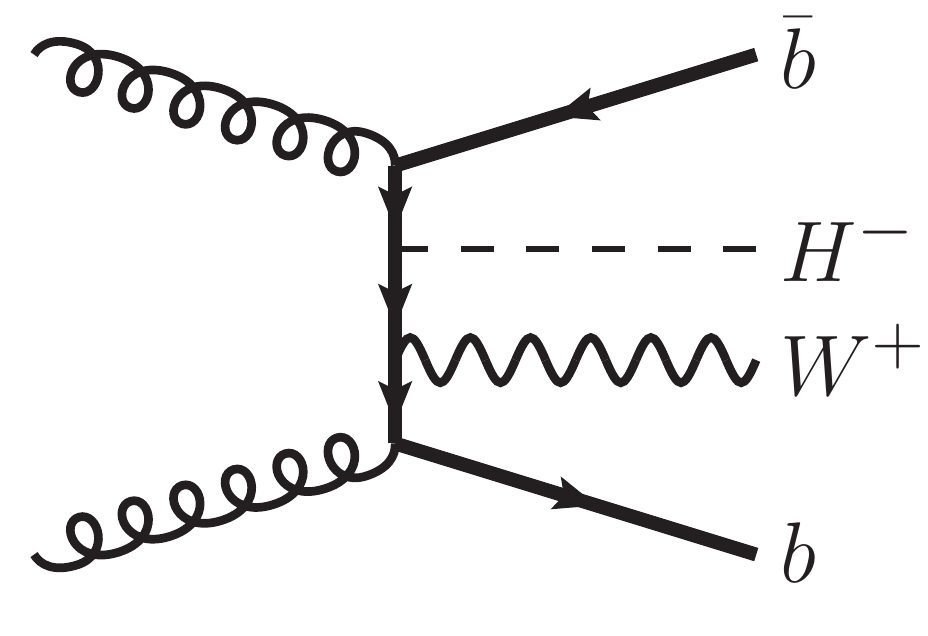} &&
    \includegraphics[clip=true, height=1.6cm]{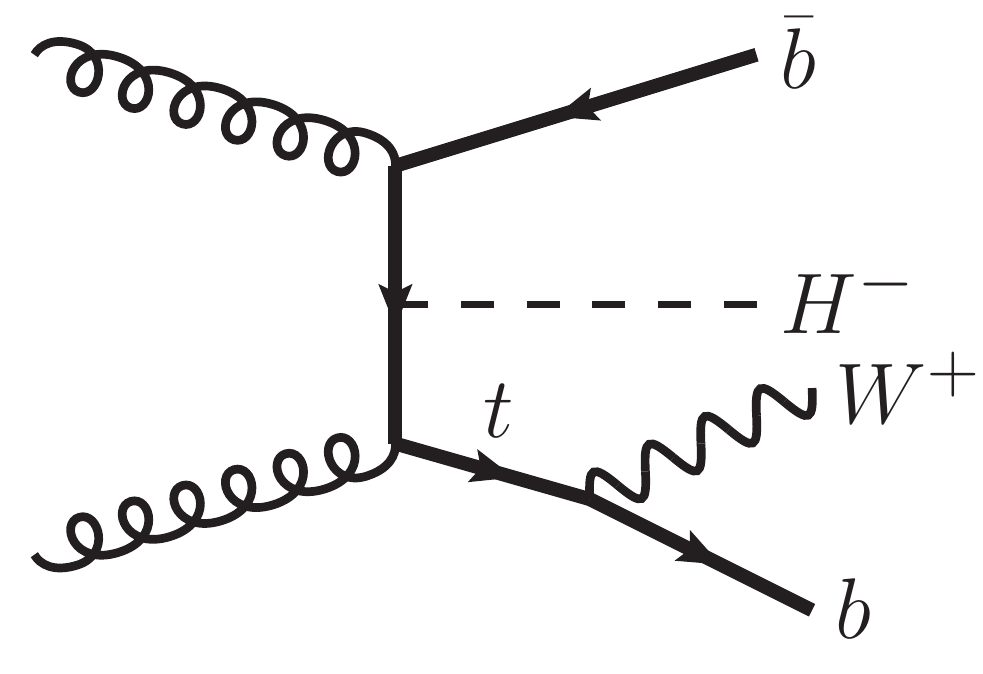} &&
    \includegraphics[clip=true, height=1.6cm]{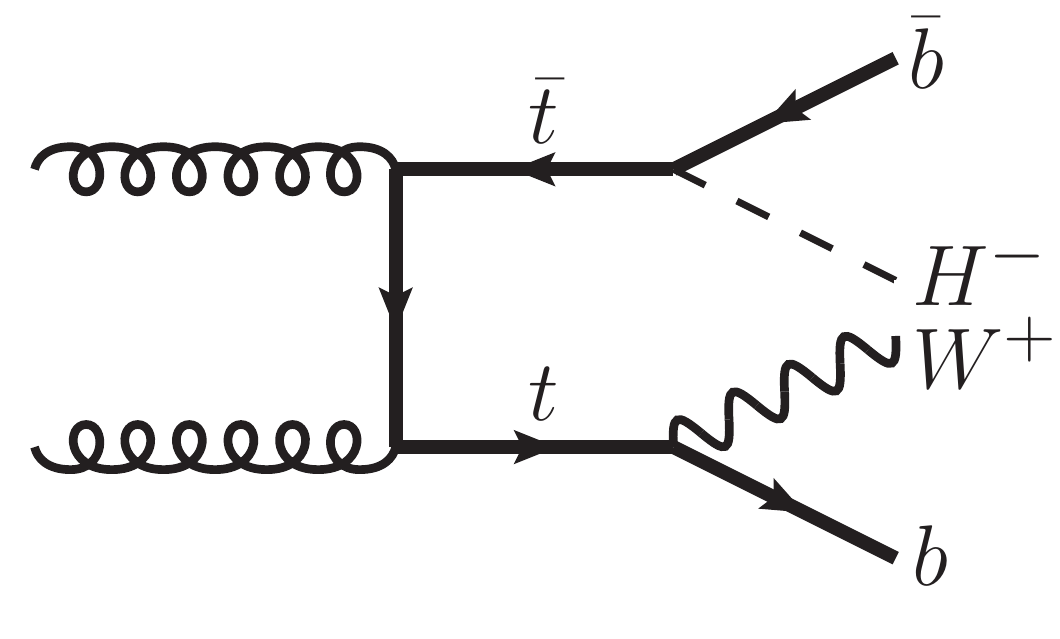} &&
    \includegraphics[clip=true, height=1.6cm]{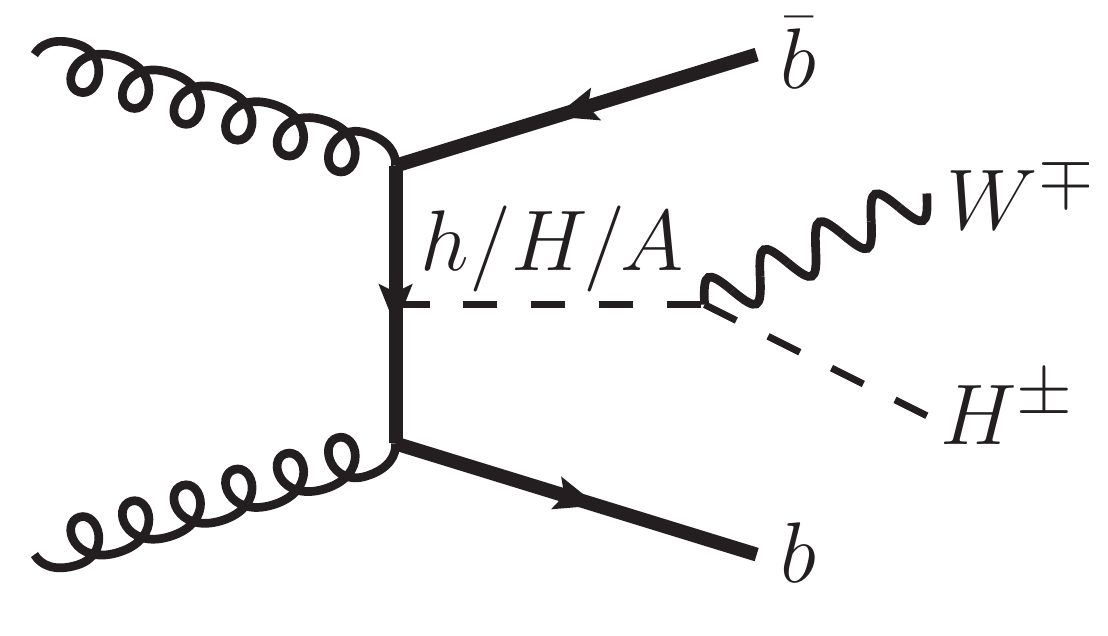}\\
     (a) && (b) && (c) && (d)
     \end{tabular}
    \caption{\label{fig:diagfull} Sample LO diagrams for the full $\hwbb$ process: (a) non-resonant top-quark contribution; (b) single-resonant top-quark contribution; (c) double-resonant top-quark contribution; (d) contribution involving neutral scalars.}
\end{figure*}

The dominant production mode for a charged Higgs boson depends on the value of its mass with 
respect to the top-quark mass, and can be classified into three categories.
Light charged Higgs scenarios are defined by Higgs-boson masses 
smaller than the mass of the top quark,
where the top-quark decay $t\to H^+b$ is allowed and the charged Higgs is light enough so that top-quark off-shell effects can be neglected 
(typically experimental analyses consider masses up to $m_{H^\pm}\lesssim 160\gev$). 
The cross section for the production of a light charged Higgs boson is simply given by
the product of the top-pair production cross section and the branching ratio of a top 
quark into a charged Higgs boson, see Fig.\,\ref{fig:diaglowhighmass}\,(a).
% added
The former is known up to next-to-next-to-leading order in perturbative
QCD~\cite{Czakon:2013goa} and displays a 3\% QCD scale uncertainty, while the NLO branching ratio
for $t\to H^+b$~\cite{Jezabek:1988iv,Li:1990qf,Campbell:2004ch,Czarnecki:1998qc,Chetyrkin:1999ju,Blokland:2004ye,Blokland:2005vq,Czarnecki:2010gb,Gao:2012ja,Brucherseifer:2013iv,Czarnecki:1992zm} is affected by 
a 2\% scale uncertainty due to missing higher-order QCD
contributions. Thus the theoretical accuracy for the production of a light charged Higgs boson
is at the few \% level. 
%
% removed
%Since the largest theoretical uncertainties stem from the top-pair production cross section, 
%which is currently known up to next-to-next-to-leading order in perturbative
%QCD~\cite{Czakon:2013goa}, the same theoretical accuracy can be claimed for the production of a light
%charged Higgs boson. 
The model-independent bounds on the branching ratio of a light charged Higgs boson~\cite{Brein:2007da} are transformed into limits in the $(m_{H^\pm},\tan\beta)$ plane, with $\tan\beta$ being the ratio of the vacuum expectation values of the two Higgs doublets.
Direct searches at the LHC, with a centre-of-mass energy of 7 TeV~\cite{Aad:2012tj,Aad:2012rjx,Aad:2013hla,Chatrchyan:2012vca} 
and 8 TeV~\cite{Khachatryan:2015uua,Khachatryan:2015qxa} set stringent constraints on the 
parameter space with a light charged Higgs boson. 

Heavy charged Higgs boson scenarios, on the other hand, 
correspond to charged Higgs masses larger than the top-quark mass 
(typically $m_{H^\pm}\gtrsim 200\gev$).
In this case, the dominant charged Higgs production channel is the associated
production with a top quark~\footnote{In the four-flavour scheme there is also an explicit bottom quark in the final state.}, see Fig.\,\ref{fig:diaglowhighmass}\,(b). Theoretical predictions at NLO(+PS) have been 
computed both at the inclusive and fully-differential
level in the five-flavour scheme (5FS)~\cite{Zhu:2001nt,Gao:2002is,Plehn:2002vy,Berger:2003sm,Kidonakis:2005hc,Weydert:2009vr,Klasen:2012wq,Degrande:2015vpa}
and in the four-flavour scheme (4FS)~\cite{Dittmaier:2009np,Flechl:2014wfa,Degrande:2015vpa}.
Charged Higgs searches at 7 TeV~\cite{Aad:2012tj}, 
8 TeV~\cite{Aad:2015typ,Aad:2014kga,Khachatryan:2015qxa} and 13
TeV~\cite{Aaboud:2016dig,ATLAS:2016grc,ATLAS:2016qiq} have set upper limits on 
the cross section for heavy charged Higgs production 
times branching ratio $\textrm{BR}(H^{\pm}\to \tau\nu_{\tau}$) 
for charged Higgs-boson masses ranging from 200 to 2000 GeV.
%In fact, the search in the $H^{\pm}\to tb$ channel 
%reveals an excess of events above the background-only hypothesis across 
%a wide $H^{\pm}$ mass range, 
%with up to 2.4 sigma deviation~\cite{Aad:2015typ}.

The intermediate-mass range is associated with 
charged Higgs masses close to the top-quark mass ($145\lesssim m_{H^\pm} \lesssim 200\gev$).
In this region, finite top-width effects as well as the interplay between top-quark resonant 
and non-resonant diagrams cannot be neglected. Therefore, the full process $\hwbb$ 
(with massive bottom quarks), see Fig.\,\ref{fig:diagfull}, including  
non-resonant, single-resonant and double-resonant contributions, has to be 
considered, to perform a reliable perturbative computation of the charged Higgs cross section.
The intermediate-mass range has not been searched for at the LHC to date,
mostly due to the lack of sufficiently accurate theoretical predictions, and the 
consequent shortage of specific strategies devised to increase the sensitivity to the signal. 
Despite the fact that some studies exist on the intermediate mass-range,
they are either only LO-accurate, thus affected by large theoretical 
uncertainties~\cite{Moretti:2002eu,Assamagan:2004gv,Alwall:2004xw}, or based on an incoherent
sum of the $pp\to t\bar t$ and $pp \to t H^-$ production mechanisms~\cite{Berger:2003sm,Klasen:2012wq}, 
and neglecting interferences between the two. With 
this work, where we compute the cross section for the $\hwbb$ process at NLO accuracy, 
we provide for the first time precise and theoretically consistent predictions in the 
intermediate-mass range, which are an essential ingredient for $H^\pm$ searches at Run II of the LHC. 
We leave to further work in collaboration with our experimental colleagues 
to devise appropriate cuts and selection strategies that would maximise the sensitivity 
to this particular mass range.
Despite the fact that exclusion bounds from flavor physics for a type-II Higgs doublet model are now very strong and exclude charged Higgs boson lighter
  than 380 GeV\footnote{See e.g. refs.~\cite{Deschamps:2009rh,Broggio:2014mna,Das:2015qva,Barranco:2016njc}}, the intermediate-mass region is not excluded for type-I models nor for models
  that embed the 2HDM-II at tree-level. 
Indeed, the intermediate-mass range has recently gained extra attention
in the model-building community. For example, supersymmetric scenarios where the heavy Higgs boson of the spectrum has a mass of 125 GeV and the light Higgs can possibly act as 
a mediator to the dark-matter sector lead to a charged Higgs-boson mass similar to the top-quark mass~\cite{Bechtle:2016kui, Profumo:2016zxo}. 
In fact, at tree-level, the Higgs-fermion Yukawa couplings of the MSSM and p-MSSM follow the 2HDM-II pattern. However, when radiative corrections are 
included,
the Yukawa couplings are modified by supersymmetry-breaking effects, thus leading to a different phenomenology. It is important to notice that such modifications of the Yukawa coupling can be included in our calculation, as it is explicitly spelled out in the following.

Our computation employs a chain of automatic tools in the
\amc+\nlo\ framework~\cite{Alwall:2014hca, Degrande:2014vpa},
developed to study the phenomenology of new physics models at NLO accuracy. In
this framework, \nlo{} automatically computes the $R_2$ rational terms and the ultraviolet
counterterms used in the virtual amplitudes, and relies internally upon {\sc FeynRules}~\cite{Alloul:2013bka}~and {\sc FeynArts}~\cite{Hahn:2000kx}. The one-loop matrix elements are computed using the
\ml~module~\cite{Hirschi:2011pa}, which employs {\sc CutTools}~\cite{Ossola:2007ax} and
{\sc Ninja}~\cite{Mastrolia:2012bu,Peraro:2014cba,Hirschi:2016mdz} for loop reduction at the integrand level and IREGI~\cite{iregi} for tensor integral reduction.
All methods are complemented by an in-house implementation of the
{\sc OpenLoops}~\cite{Cascioli:2011va} algorithm. For the
factorisation of the IR poles in the real-emission phase-space
integrals, the resonance-aware {\sc MadFKS}~\cite{Frederix:2009yq,Frederix:2016rdc} module is
used. 

We work in the four-flavour scheme, where the bottom-quark mass regulates
any soft or collinear divergence related to final-state bottom-quark emissions, making it possible to compute the
total cross section without having to impose artificial cuts on the final state particles. In a 5FS version 
of this computation ($b\bar b\to H^\pm W^\mp$), \mbox{non-,} single- and double-resonant contributions are included at different accuracies. In particular 
the double-resonant contributions only enter at NNLO (and beyond). Even in that case, these contributions would be effectively included 
only at lowest order, hampering the formal accuracy of the computation in the region $m_{H^\pm}<m_t$, where they are dominant. On the 
other hand, in our 4FS calculation all contributions are included at NLO accuracy. Moreover, the 4FS has been shown 
to provide reliable predictions
for the heavy-Higgs case~\cite{Dittmaier:2009np,Degrande:2015vpa}, without being spoiled by large logarithms.
For consistency, we use the four-flavour set of the PDF4LHC15 parton 
distributions~\cite{Butterworth:2015oua, Ball:2014uwa, Harland-Lang:2014zoa, Dulat:2015mca}, 
and the corresponding running of $\alphas$ with $\alphas(m_Z)=0.1126$.

The identification of the hard scales in a complex process, such as the one at hand, is not necessarily a trivial task.
One has to bear in mind, however, that in the intermediate region 
it is desirable to have a matching to the scale in the $pp\to t\bar t$ cross section 
for light charged Higgs masses, where the natural choice is of the order of the 
top-quark mass (or below~\cite{Czakon:2016dgf}), and for larger masses to the scale in the heavy charged Higgs cross section, 
where the scale $\mu = (m_t + m_{H^\pm} + m_b)/3$ is typically applied in 4FS computations.
We therefore fix our renormalisation and factorisation scales ($\mu_r$ and $\mu_f$) to $\mu=125\, \gev$, which
matches the numerical value used for the heavy charged Higgs production at 
$m_{H^\pm}=200\,\gev$, while it satisfies the requirement of being in between $m_t/2$ and $m_t$ 
for the light charged-Higgs case. 

The top-quark mass and Yukawa coupling are renormalized on-shell, while we use a hybrid scheme for the bottom-quark mass: kinematical bottom-quark masses are treated with an on-shell renormalization, but the $\overline{\rm{MS}}$ renormalisation scheme is employed for the bottom-quark Yukawa coupling.
For the numerical values we follow the recommendations of the LHC Higgs Cross Section Working Group~\cite{Denner:2047636}, which implies
$m_t^{\rm{OS}} = 172.5\,\gev$ and $m_b^{\rm{OS}}=4.92\, \gev$ for the on-shell masses. Using the four-loop conversion~\cite{Marquard:2015qpa}~and running, this corresponds 
to the $\overline{\rm{MS}}$ bottom mass $m_b(m_b)\simeq 4.18\, \gev$ and $m_b(\mu) \simeq 2.81\,\gev$, respectively.
For the computation of scale variations starting from $m_b(\mu)$, a two-loop running is employed.

Since the $\hwbb$ process involves resonant
top-quark contributions, the width of the top quark has to
be included in the computation without spoiling gauge invariance. This is achieved by employing the complex-mass scheme~\cite{Denner:1999gp,Denner:2005fg}, where 
the top-quark mass (and Yukawa coupling) are regarded as complex parameters.
For a given charged Higgs mass and $\tan \beta$, we compute the corresponding top-quark width at the
same perturbative order in $\alphas$ as the cross section. The charged Higgs boson and
the $W$ boson are kept on-shell.

Compared to calculations of similar complexity (e.g. the $pp\to W^+ W^- b\bar b$ process in the 4FS~\cite{Frederix:2013gra, Cascioli:2013wga}), 
the technical challenges of this process lie in the interplay between the non-, single- and double-resonant contributions, which can have a different 
hierarchy depending on $m_{H^\pm}$. On top of this, the cross section receives contributions with different powers of the bottom-quark Yukawa coupling, and therefore its running cannot 
be accounted for through an overall factor. Unlike in previous computations~\cite{Wiesemann:2014ioa,Degrande:2015vpa} these contributions, 
including scale variations, are computed simultaneously.

Among the various Feynman diagrams contributing to the $\hwbb$ process,
some include the neutral Higgs states of the 2HDM ($h$, $H$, $A$) and their coupling to bottom quarks, see Fig.\,\ref{fig:diagfull}\,(d).
We refrain from including these contributions in our computation at NLO, but briefly comment
on the size of their effects below. To be able to make quantitive statements we must 
make some assumptions regarding the 2HDM parameters. We use the 
so-called ``alignment'' region ($\cos(\beta-\alpha)\simeq 0$, with $\alpha$ the mixing angle of the two CP even scalars), 
where the $125$\,GeV Higgs boson discovered at 
the LHC corresponds to the light scalar $h$~\cite{Craig:2015jba}.\footnote{
    Models where the heavy Higgs corresponds to the particle discovered at the LHC, such as those from Ref.~\cite{Profumo:2016zxo}, tend to prefer small or moderate 
    values of $\tan \beta$, which greatly suppress this kind of diagrams.}
In principle, $m_A$ and $m_H$ can be chosen such that the $H$ and $A$ states may become resonant. In practice, if this choice is made, one is {\it de facto} considering the simpler process 
$pp\to H/A b \bar b$, with $H/A \to H^\pm W^\mp$ decay. Therefore, we will not consider this case here.
We have verified that the impact of the neutral Higgs states is completely 
negligible for small $\tan \beta$. At large
$\tan \beta$ ($\tan \beta = 30$), we found at most $-7\%$ impact on 
the LO cross section for $m_{H^\pm} > 180~\gev$ 
in the configuration $m_H = m_A \simeq m_{H^\pm} - 45~\gev$. 
For other values of $m_{H^\pm}$ and for 
heavier neutral Higgses the effect is smaller. 
Lighter neutral Higgses are strongly disfavoured by 
EW precision fits~\cite{Denner:1991ie,Chankowski:1999ta,LopezVal:2012zb} and direct searches.
We thus reckon that our choice of not including contributions from neutral Higgs bosons 
is justified, as their small impact can be included separately and off-line at LO without 
hampering the accuracy of our NLO results presented below.

We now present our results for the total cross section of the $pp \to H^+ W^- b \bar b$ process (the charge-conjugated process has the same total cross section)
at NLO QCD, at the 13 TeV LHC. We consider three different values of the $\tan\beta$ parameter, $\tan \beta = 1, 8, 30$. The total cross sections
at LO and NLO accuracy in the range $m_{H^\pm}/{{\rm GeV}}\in[145, 200]$ are given in Tab.~\ref{tab:xsec}, together with the NLO $K$-factors, defined as
the ratio $K=\sigma_{\rm NLO}/\sigma_{\rm LO}$.
Next to the total cross sections, we quote the scale and PDF uncertainties.
Scale uncertainties are computed by varying independently the renormalisation and factorisation scales in the range $\mu_r, \mu_f \in [ \mu/2 , 2 \mu]$ (albeit keeping the scale in the computation of the top-quark width fixed to the central value), while
for PDF uncertainties we follow the PDF4LHC15 procedure \cite{Butterworth:2015oua}. NLO corrections are large; they increase the central value of the total cross section by $50\%-60\%$, with only a very mild dependence on the charged Higgs-boson mass and $\tan\beta$ value, and 
significantly reduce the scale dependence with respect to LO. More precisely, NLO scale uncertainties range between $8\%-13\%$ ($10\%-17\%$) for $m_{H^\pm}<m_t$ ($m_{H^\pm}>m_t$).
In both cases, the large-$\tan\beta$ ($\sigma \sim y_b^2$) scenario features larger scale uncertainties than the small-$\tan\beta$ ($\sigma\sim y_t^2$) one, because of the additional
$\mu_r$-dependence introduced by the running of the bottom-quark Yukawa coupling.\\

\begin{figure}
    \centering
    \includegraphics[width=0.5\textwidth]{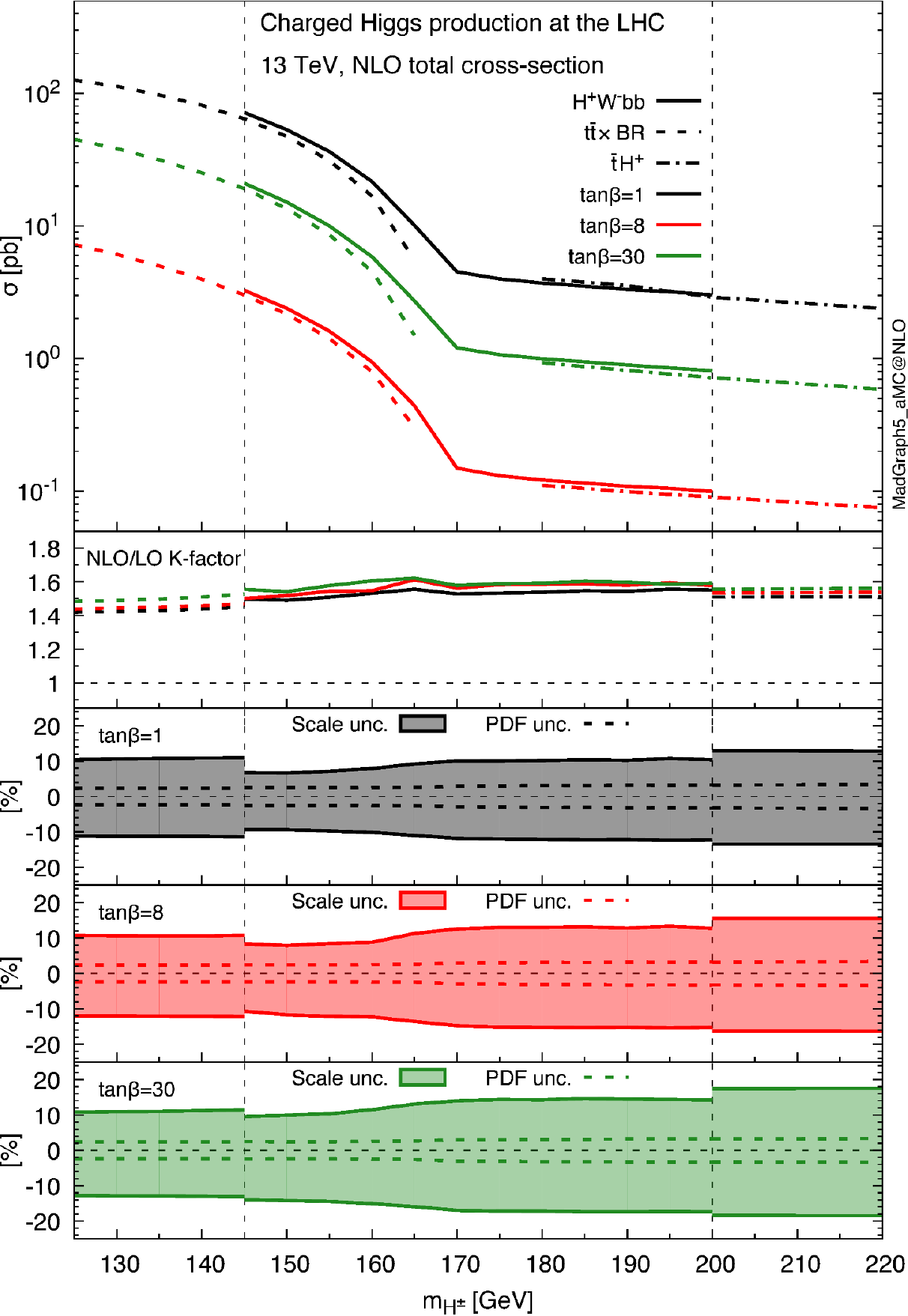}
    \caption{\label{fig:nloxsect} NLO total cross sections, $K$-factors and uncertainties for charged Higgs boson production at the 13 TeV LHC.}
\end{figure}

\begin{table*}[ht!] % table* makes the table and caption span the two columns
\centering
\footnotesize
\begin{tabular}{p{0.15cm}@{$\qquad$}rrp{0.15cm}@{$\qquad$}rrp{0.15cm}@{$\qquad$}rrp{0.15cm}}
\hline
$m_{H^\pm}$ & \multicolumn{3}{c@{$\qquad$}}{$\tan\beta=1$} & \multicolumn{3}{c@{$\quad$}}{$\tan\beta=8$} & \multicolumn{3}{c}{$\tan\beta=30$}\vspace{-1pt}\\
$[\gev]$   &   \multicolumn{1}{c}{$\sigma_{\rm LO}$}  &   \multicolumn{1}{c}{$\sigma_{\rm NLO}$}   &   $K$   &
			 \multicolumn{1}{c}{$\sigma_{\rm LO}$}   &   \multicolumn{1}{c}{$\sigma_{\rm NLO}$}   &   $K$   &
			 \multicolumn{1}{c}{$\sigma_{\rm LO}$}   &   \multicolumn{1}{c}{$\sigma_{\rm NLO}$}   &   $K$   \\
\hline
        $145$   &   $  47.8 ^{+31} _{-22} \pm 2.4 $   &   $  71.6 ^{+ 7} _{-9} \pm 2.4 $   &   $1.50$   &   $  2.17 ^{+39} _{-26} \pm 2.4 $   &   $  3.26 ^{+ 8} _{-11} \pm 2.4 $   &   $1.50$   &   $  13.5 ^{+46} _{-29} \pm 2.4 $   &   $  21.0 ^{+10} _{-14} \pm 2.5 $   &   $1.55$   \\
        $150$   &   $  35.7 ^{+31} _{-22} \pm 2.4 $   &   $  53.1 ^{+ 7} _{-9} \pm 2.4 $   &   $1.49$   &   $  1.57 ^{+39} _{-26} \pm 2.4 $   &   $  2.38 ^{+ 8} _{-12} \pm 2.4 $   &   $1.52$   &   $  9.81 ^{+46} _{-29} \pm 2.4 $   &   $  15.1 ^{+10} _{-14} \pm 2.4 $   &   $1.54$   \\
        $155$   &   $  24.1 ^{+31} _{-22} \pm 2.4 $   &   $  36.3 ^{+ 7} _{-10} \pm 2.4 $   &   $1.51$   &   $  1.04 ^{+39} _{-26} \pm 2.4 $   &   $  1.61 ^{+ 8} _{-12} \pm 2.4 $   &   $1.54$   &   $  6.34 ^{+46} _{-29} \pm 2.4 $   &   $  9.99 ^{+10} _{-14} \pm 2.4 $   &   $1.58$   \\
        $160$   &   $  14.1 ^{+31} _{-22} \pm 2.5 $   &   $  21.6 ^{+ 8} _{-10} \pm 2.5 $   &   $1.53$   &   $ 0.609 ^{+39} _{-26} \pm 2.4 $   &   $ 0.943 ^{+ 9} _{-12} \pm 2.5 $   &   $1.55$   &   $  3.64 ^{+47} _{-29} \pm 2.5 $   &   $  5.85 ^{+11} _{-15} \pm 2.5 $   &   $1.60$   \\
        $165$   &   $  6.50 ^{+32} _{-23} \pm 2.6 $   &   $  10.1 ^{+ 9} _{-11} \pm 2.6 $   &   $1.56$   &   $ 0.274 ^{+40} _{-26} \pm 2.5 $   &   $ 0.442 ^{+11} _{-14} \pm 2.5 $   &   $1.61$   &   $  1.68 ^{+48} _{-30} \pm 2.6 $   &   $  2.72 ^{+13} _{-16} \pm 2.6 $   &   $1.62$   \\
        $170$   &   $  2.95 ^{+34} _{-23} \pm 2.9 $   &   $  4.51 ^{+10} _{-12} \pm 3.0 $   &   $1.53$   &   $ 0.095 ^{+43} _{-27} \pm 2.9 $   &   $ 0.149 ^{+13} _{-15} \pm 3.0 $   &   $1.56$   &   $ 0.763 ^{+50} _{-31} \pm 3.0 $   &   $  1.20 ^{+14} _{-17} \pm 3.0 $   &   $1.58$   \\
        $175$   &   $  2.60 ^{+34} _{-24} \pm 3.0 $   &   $  3.98 ^{+10} _{-12} \pm 3.0 $   &   $1.53$   &   $ 0.083 ^{+43} _{-28} \pm 3.0 $   &   $ 0.131 ^{+13} _{-15} \pm 3.0 $   &   $1.58$   &   $ 0.674 ^{+51} _{-31} \pm 3.1 $   &   $  1.07 ^{+14} _{-17} \pm 3.1 $   &   $1.59$   \\
        $180$   &   $  2.41 ^{+34} _{-24} \pm 3.1 $   &   $  3.71 ^{+10} _{-12} \pm 3.1 $   &   $1.54$   &   $ 0.077 ^{+44} _{-28} \pm 3.1 $   &   $ 0.121 ^{+13} _{-15} \pm 3.2 $   &   $1.59$   &   $ 0.627 ^{+51} _{-31} \pm 3.1 $   &   $ 0.998 ^{+14} _{-17} \pm 3.2 $   &   $1.59$   \\
        $185$   &   $  2.27 ^{+35} _{-24} \pm 3.1 $   &   $  3.51 ^{+10} _{-12} \pm 3.1 $   &   $1.55$   &   $ 0.073 ^{+44} _{-28} \pm 3.1 $   &   $ 0.115 ^{+13} _{-15} \pm 3.1 $   &   $1.59$   &   $ 0.591 ^{+51} _{-31} \pm 3.2 $   &   $ 0.947 ^{+15} _{-17} \pm 3.2 $   &   $1.60$   \\
        $190$   &   $  2.15 ^{+35} _{-24} \pm 3.1 $   &   $  3.32 ^{+10} _{-12} \pm 3.2 $   &   $1.54$   &   $ 0.069 ^{+44} _{-28} \pm 3.2 $   &   $ 0.109 ^{+13} _{-15} \pm 3.2 $   &   $1.58$   &   $ 0.561 ^{+51} _{-31} \pm 3.2 $   &   $ 0.896 ^{+14} _{-17} \pm 3.3 $   &   $1.60$   \\
        $195$   &   $  2.05 ^{+35} _{-24} \pm 3.2 $   &   $  3.18 ^{+11} _{-12} \pm 3.2 $   &   $1.56$   &   $ 0.066 ^{+44} _{-28} \pm 3.2 $   &   $ 0.105 ^{+13} _{-15} \pm 3.2 $   &   $1.60$   &   $ 0.536 ^{+52} _{-32} \pm 3.2 $   &   $ 0.850 ^{+14} _{-17} \pm 3.2 $   &   $1.59$   \\
        $200$   &   $  1.95 ^{+35} _{-24} \pm 3.2 $   &   $  3.02 ^{+10} _{-12} \pm 3.3 $   &   $1.55$   &   $ 0.063 ^{+44} _{-28} \pm 3.2 $   &   $ 0.100 ^{+13} _{-15} \pm 3.3 $   &   $1.58$   &   $ 0.510 ^{+52} _{-32} \pm 3.3 $   &   $ 0.812 ^{+14} _{-17} \pm 3.3 $   &   $1.59$   \\

\hline
\end{tabular}
\caption{\label{tab:xsec} LO and NLO total cross sections (in pb) and $K$-factors for the $pp\to H^+ W^- b \bar b$ process, for $\tan\beta=1,8,30$ at the 13 TeV LHC. The first quoted uncertainties
are from scale variations, the second from PDFs (both in per cent of the total cross section). The statistical uncertainty from the numerical phase-space integration is of the order of 1\% or below.}
\end{table*}
Further details on the behaviour of the scale uncertainties can be inferred from Fig.\,\ref{fig:nloxsect},
where we compare our intermediate-mass range results to dedicated predictions
for light and heavy charged Higgs production.  
The input parameters have been chosen consistently across all the mass range, in particular all cross sections are computed in the 4FS, the central scale for low-mass range is also set to $\mu=125~\gev$, while the scale $\mu = (m_t + m_{H^\pm} + m_b)/3$ is used for the heavy charged Higgs case.
The central predictions in the main frame develop a prominent structure 
with a kink at the threshold $m_{H^\pm}\simeq m_t-m_b$.
The effect of the single-resonant contributions ($pp\to t W^-$ and $pp \to \bar t H^+$) is visible when comparing our results in the intermediate-mass range with the low-mass prediction. Indeed, 
the single-resonant contributions 
are missing in the low-mass prediction and amount to $10\%-15\%$ of the $pp \to t\bar t$ cross section depending on the specific
value of $\tan\beta$. In contrast,
looking at the matching of the intermediate-mass predictions to the heavy charged Higgs cross section, we observe a $5\%-10\%$ gap for $\tan\beta=8$ and $\tan\beta=30$, while there is essentially no gap for 
$\tan\beta=1$. Such a gap originates from the non-resonant part of the $\hwbb$ amplitude, which, because of the chiral structure of the $H^+ tb$ and $Wtb$ vertices, is enhanced 
(suppressed) for large (small) values of $\tan \beta$. At 145 and 200
GeV, the size of the scale uncertainty in the intermediate region and
the side-bands is slightly different. These discontinuities are
related to missing subleading terms in the predictions used in the low
and high-mass regions, i.e.~mostly single-resonant and non-resonant,
respectively, although it is difficult to pin down exactly the origin 
of the discontinuities
because of the non-trivial separation of these contributions beyond
leading order.
Finally, we note that the $K$-factor in the intermediate region interpolates very well the ones in the low and high-mass range.

We now discuss how to generalise our results at a single $\tan\beta$ value in order to obtain the charged Higgs boson cross section in the 
intermediate-mass range for any value of $\tan\beta$ or in a type-I 2HDM by means of reweighting. As discussed in 
Ref.~\cite{Degrande:2015vpa}, the cross section for
charged Higgs production receives
contributions proportional to $y_b^2$, $y_t^2$ and $y_b y_t$. In a type-II 2HDM, while the $y_b y_t$ contribution 
does not depend on $\tan \beta$, the $y_b^2$ and $y_t^2$ ones scale as $\tan\beta^2$ and $1/\tan\beta^2$, respectively. Conversely,
in a type-I 2HDM, all contributions (and therefore the total cross section) scale as $1/\tan\beta^2$. We point out that a naive reweighting,
such as the one proposed in Ref.~\cite{Degrande:2015vpa} for a heavy charged Higgs boson, is bound to fail in our case, since it 
will miss effects due to the $\tan\beta$ dependence of the top width. We verified that, if the top-width 
dependence is included as an overall factor, we are able to reproduce our $\tan\beta=1$ and $\tan\beta=30$ NLO cross sections and uncertainties starting from the numbers at $\tan\beta=8$ with an accuracy of 1\% or better,
using the relation (the dependence on $m_{H^\pm}$ is understood)%~\footnote{Cross sections for other scenarios than a type-II 2HDM can be computed by trivially extending eq.~(1) according to the scaling of the various contributions in a given scenario. For example, in a type-I 2HDM, all three contributions
% to the cross section would scale as $(\tan \beta/\tan \beta')^2$.}
\begin{equation}
    \sigma^{\rm{t-II}}(\tan\beta') = \left[\left(\frac{\tan\beta'}{\tan\beta}\right)^2 \sigma^{\rm{t-II}}_{y_b^2}(\tan\beta)
                              +\sigma^{\rm{t-II}}_{y_b y_t} (\tan\beta)
                              +\left(\frac{\tan\beta}{\tan\beta'}\right)^2 \sigma^{\rm{t-II}}_{y_t^2}(\tan\beta)
                         \right] \times \left(\frac{\Gamma_t(\tan\beta)}{\Gamma_t(\tan\beta')}\right)^2\,.
\end{equation}
This also shows that effects due to the width-dependent complex phase
of $y_t$ are very small.
Concerning how to extend our results in a type-I 2HDM, we first point out that for $\tan \beta=1$, the cross-section is identical to the type-II case. Then,
the cross-section for any other value of $\tan \beta$ can be simply obtained as
\begin{equation}
    \sigma^{\rm{t-I}}(\tan\beta') =\frac{\sigma^{\rm{t-I}}(\tan\beta=1)}{(\tan\beta')^2}  \times \left(\frac{\Gamma_t(\tan\beta)}{\Gamma_t(\tan\beta')}\right)^2\,.
\end{equation}
Exploiting Eqs.~(1) and ~(2) we produced cross section tables 
for $\tan\beta\in [0.1, 60]$, both for a type-II and a type-I 2HDM, which are publicly available~\footnote{\url{https://twiki.cern.ch/twiki/bin/view/LHCPhysics/LHCHXSWGMSSMCharged\#Intermediate_mass_145_200_GeV_ch}}. Finally, Eq.~(1) can also be used to include the dominant supersymmetric corrections, in particular those which modify the relation between the bottom-quark mass and its Yukawa coupling. These corrections are enhanced at large $\tan\beta$ and can be resummed to all orders by modifying the bottom-quark Yukawa coupling~\cite{Dittmaier:2009np}.

In conclusion, we have presented predictions for the production of an intermediate-mass charged Higgs boson. While we have focused on the case of a type-II 2HDM, our results can be easily extended
to other scenarios, such as a type-I 2HDM or supersymmetry. For the first
time theoretically consistent predictions at NLO QCD accuracy have been made available in this mass range. To this end, we have studied the 
$\hwbb$ process in the complex-mass scheme, including finite top-width effects and contributions with resonant top quarks. Our results provide a reliable interpolation of low- and high-mass regions and
make it possible to finally extend direct searches for charged Higgs bosons to the $m_{H^\pm}\sim m_t$ region,
so far unexplored by LHC experiments. 
The central value of the NLO total cross section is well-approximated
by a factor of about $1.5-1.6$ times the LO cross section, with only a
very mild dependence on the charged Higgs mass and $\tan\beta$. 
% added
The results presented in  paper constitute an important step in filling a 
gap in the available theoretical predictions for charged Higgs boson production 
at next-to-leading order in QCD. Current results could be further improved by 
including model-dependent sub-leading contributions that may become
dominant in case of large width of heavy neutral Higgses, and by considering 
differential distributions.
We leave it to future work to study if this factorisation of the NLO
corrections also holds at the same level for differential
distributions, employing modern techniques developed to take into account internal resonances when matching NLO computations with parton showers~\cite{Jezo:2015aia, Frederix:2016rdc, Jezo:2016ujg}.

\section*{Acknownledgements}

We are indebted to Fabio Maltoni for his constant encouragement and his valuable suggestions.
We would like to thank the LHC Higgs Cross Section Working Group for providing 
the motivation to perform this calculation. We are especially grateful to Martin Flechl, Stephen Sekula and Michael Spira for their comments on the manuscript.
MZ would like to thank Pietro Slavich and Tim Stefaniak for discussions.
CD is a Durham International Junior Research Fellow; the work of RF is supported by the Alexander von
Humboldt Foundation, in the framework of the Sofja Kovaleskaja Award Project
``Event Simulation for the Large Hadron Collider at High Precision'';
the work of VH is supported by the Swiss National Science Foundation (SNF) with grant PBELP2 146525;
the work of MW is supported by the Swiss National Science
Foundation (SNF) under contract 200020-141360; and the work of MZ is supported by the
European Union's Horizon 2020 research and innovation
programme under the Marie Sklodovska-Curie grant agreement No 660171 and in part by 
the ILP LABEX (ANR-10-LABX-63), in turn supported by French state funds
managed by the ANR within the ``Investissements d'Avenir'' programme under
reference ANR-11-IDEX-0004-02.

\section*{References}

\bibliography{cHint}

\end{document}